\documentclass[conference]{IEEEtran}
\IEEEoverridecommandlockouts
\usepackage{cite}
\usepackage{amsmath,amssymb,amsfonts}
\usepackage{algorithmic}
\usepackage{graphicx}
\usepackage{textcomp}
\usepackage{xcolor}
\def\BibTeX{{\rm B\kern-.05em{\sc i\kern-.025em b}\kern-.08em
		T\kern-.1667em\lower.7ex\hbox{E}\kern-.125emX}}
\def\ie{{i.e., }}

\usepackage{makecell}
\usepackage{pifont}
\usepackage{ragged2e}
\usepackage{floatrow}
\usepackage{multirow}
\usepackage{amsthm}
\usepackage[utf8]{inputenc}
\usepackage{soul}            
\usepackage[english]{babel}
\usepackage{graphicx}
\usepackage{tikz}
\usepackage[utf8]{inputenc}
\theoremstyle{remark}

\newtheorem{Proposition}{Proposition}

\usepackage{multicol}
\usepackage{epsfig}
\usepackage{epstopdf}
\usepackage{float}
\usepackage{perpage}
\MakeSorted{figure}
\MakeSorted{table}
\usepackage{array}
\usepackage{stfloats}
\usepackage{graphicx}
\usepackage{amsfonts}
\usepackage{amssymb}

\usepackage{enumitem}
\usepackage{verbatim}
\usepackage[english]{babel}
\usepackage[utf8]{inputenc}
\usepackage[linesnumbered,ruled,vlined]{algorithm2e}
\usepackage{cite}
\DeclareMathAlphabet\mathbfcal{OMS}{cmsy}{b}{n}
\ifCLASSINFOpdf
\else
\fi
\usepackage{array}
\usepackage{fixltx2e}

\usepackage[bf,small]{caption}
\usepackage[font=footnotesize,labelfont=bf, figurename=Fig.]{caption} 
\usepackage{subcaption}
\usepackage{fix-cm}


\usepackage{etoolbox} 
\newtoggle{SINGLE_COL}
\toggletrue{SINGLE_COL}     

\newtoggle{BIG_EQUATION}
\togglefalse{BIG_EQUATION}  

\usepackage{mathtools}

\newcommand{\Imat}{\mathbf{I}}

\newcommand{\Rmat}{\mathbf{R}}

\newcommand{\hv}{\mathbf{h}}
\newcommand{\nv}{\mathbf{n}}

\newcommand{\yv}{\mathbf{y}}

\newcommand{\fv}{\mathbf{f}}
\newcommand{\vv}{\mathbf{v}}
\newcommand{\nonum}{\nonumber}

\newcommand{\Deltamat}{\boldsymbol{\Delta}}
\newcommand{\Lambdamat}{\boldsymbol{\Lambda}}

\newcommand{\E}{\mathbb{E}}
\newcommand{\Tr}{\mbox{Tr}}

\newfloatcommand{capbtabbox}{table}[][\FBwidth]
\DeclareCaptionLabelFormat{andtable}{#1~#2  \&  \tablename~\thetable}
\newcommand{\beeq}{\begin{align}}
	\newcommand{\eeqs}{\end{align}}
\begin{document}

	\title{ 
	Is Downlink Training Necessary for User-Centric
	Cell-Free RSMA Systems With Mobile Users? \\
}
\author{\IEEEauthorblockN{Ravi Kiran Palla, Dheeraj Naidu Amudala and Rohit Budhiraja, \textit{IIT Kanpur,} India}
	\IEEEauthorblockA{\{prkiran, dheeraja, rohitbr\}@iitk.ac.in}
	
}
\maketitle
\begin{abstract} 
	We study the spectral efficiency (SE) of a rate-splitting multiple access (RSMA) enabled multi-clustered cell-free (CF) massive multiple-input multiple-output (mMIMO) system. The access points (APs) in each cluster serve mobile user equipments (UEs) by employing RSMA. The UEs employ successive interference cancellation  to decode their data. This work emphasizes the role of downlink (DL) pilots in realizing RSMA benefits in practical CF systems with spatially-correlated Rician channels which observe random phase shifts, pilot contamination, and channel aging due to UE mobility. We numerically show that DL pilots are required for RSMA in user-centric CF mMIMO systems with channel aging to outperform spatial division multiple access. We show that the degraded channel quality due to higher UE velocity and longer resource block lengths significantly reduces the RSMA SE. Increasing the number of clusters can compensate for the SE loss. 
\end{abstract}
\IEEEpeerreviewmaketitle
\vspace{-0.3cm}
\section{Introduction}
User-centric cell-free (CF) massive MIMO systems efficiently mitigate the tight synchronization requirements and high fronthaul load of conventional CF systems~\cite{Aboulfotouh2025,Singh2023,Hu2024,Jiang2024}. Each user equipment (UE) in a user-centric CF mMIMO system connects to a selected subset of nearby APs. This enables the system to not only support a higher number of UEs,  but also reduce the fronthaul load~\cite{Aboulfotouh2025,Singh2023,Hu2024,Jiang2024}. The authors in~\cite{Aboulfotouh2025,Singh2023} designed pilot assignment strategies to reduce the pilot overhead, and increase the spectral efficiency (SE). Hu \textit{et al.} in~\cite{Hu2024} and Jiang \textit{et al.} in~\cite{Jiang2024} further enhanced the SE by optimally selecting  most suitable APs to serve each UE. The multi-cluster interference (MCI), which arises from APs serving UEs in other clusters, however, remains a major challenge in user-centric CF systems. 

The aforementioned user-centric CF studies have employed space division multiple access (SDMA), where multiple UEs are spatially multiplexed on the same spectral resource. The SDMA, however,  may not perform well, when UEs experience MCI~\cite{Flores2023}, which is specific to CF systems. Rate splitting multiple access (RSMA) splits a user message into common and private parts, and mitigates MCI by partially decoding it using successive interference cancellation (SIC). Several CF mMIMO works have recently investigated the use of RSMA~\cite{ Zheng2024, Zheng2025,Flores2023,Ziheng2025}. For example, for \textit{conventional} CF mMIMO RSMA systems,  Zheng \textit{et al.} in~\cite{Zheng2024} designed a precoder to mitigate the multi user interference, and the authors in~\cite{Zheng2025} studied the impact of imperfect channel state information (CSI). For \textit{user-centric} CF mMIMO systems, very few studies have investigated the  RSMA benefits  \cite{Flores2023,Ziheng2025}. Ziheng \textit{et al.} in \cite{Ziheng2025} designed a two-layer RSMA framework to handle the MCI. The authors in~\cite{Flores2023} employed RSMA to enhance the SE.

The above CF mMIMO studies in \cite{Zheng2025,Flores2023} assume a block fading model, where channel remains constant in the entire resource block. Future sixth generation  networks are being designed to support UEs speed up to $500$ km/h~\cite{Enescu2008}, with their channels varying within a resource block. The CSI estimated by APs in the uplink for such UEs quickly become outdated. This phenomenon, known as channel aging, can significantly degrade the SE of CF mMIMO systems~\cite{Zheng2024,Chen2025}. 

The UEs in the existing DL CF SDMA literature either use statistical CSI to decode data~{\cite{Zhang2024}}, or estimate the instantaneous CSI by using the DL pilots transmitted by the APs \cite{Yao2024}. Statistical CSI is preferable in co-located mMIMO systems where channel hardens \cite{Yao2024}. Instantaneous CSI is preferable in CF mMIMO systems wherein channel does not harden \cite{Yao2024}. In the DL CF RSMA systems considered herein, UEs need CSI to perform SIC of common message before decoding their private message. Channel aging considered herein also makes the   estimated DL CSI outdated. The \textit{instantaneous and aged} DL CSI  will  not only degrade the SIC quality of common message, but also impact the mitigation of MCI, which UEs observe in the \textit{multi-clustered} user-centric CF mMIMO system considered herein. In this work, to the best of our knowledge, for the first time, we investigate the effect of \textit{instantaneous and aged} DL CSI  estimated using DL pilots on the twin aspects of common-message SIC, and the MCI experienced by the UEs in  user-centric CF mMIMO systems.

We now list our \textbf{main contributions} which bridge the gaps in the existing literature with respect to the system architecture and channel modeling assumptions. \newline
$\bullet$ We consider a DL user-centric CF mMIMO RSMA system, wherein UEs and APs are divided into different clusters. The APs in each cluster serve their UEs by using RSMA. Each UE decodes its common and private stream data by canceling the MCI from other clusters by employing SIC.  To enhance the SIC quality and combat aging, the APs transmit precoded pilots, which the UEs use to estimate the effective DL CSI. 

$\bullet$ We numerically show the importance of DL pilots in achieving tangible RSMA gains in user-centric CF mMIMO systems with channel aging. Further, RSMA is shown to be more sensitive to the UE velocity than the SDMA. \textit{Increasing the number of clusters can help in combating the SE loss due to channel aging in scenarios with a high UE velocity.}
\vspace{-2pt}
\section{System Model}
We consider a user-centric CF mMIMO system with $M$ APs and $K$ single-antenna mobile UEs. Each AP has $N$ antennas. The UEs and APs are divided into $L$ clusters, with $\widetilde K_l$ UEs and $\widetilde M_l$ APs in the $l$th cluster, which are denoted by the sets $\mathcal{K}_l$ and $\mathcal{A}_l$, respectively. The UE-AP association is based on the dynamic cooperation clustering (DCC) protocol~\cite{demir2021}, which assumes that the sets $\mathcal{A}_l$ and $\mathcal{K}_l$ are decided a priori to communication. The APs in each cluster employ RSMA to serve their UEs by splitting the UE signals into common and private messages. The common messages of UEs in a cluster, similar to~\cite{Flores2023,Zheng2024,Zheng2025}, are encoded into a single common message. The APs in each cluster, thus, employ a single precoder for the common message, but use a different precoder for each private message~\cite{Flores2023}. Since multiple clusters transmit simultaneously, the UEs in a cluster experience MCI from the RSMA signals of other clusters. Each UE detects its common and private messages by suppressing the MCI using SIC. The mobile UEs observe time-varying channels within a resource block, which degrades the quality of estimated CSI, and consequently the RSMA gains~\cite{Chen2025}. To handle this MCI, APs send precoded DL pilots, using which the UEs estimate the effective DL CSI and use it to decode their signals via~SIC.
\vspace{-5pt}
\subsection{Transmission Protocol} 
For the DL-pilot-aided RSMA, the resource block of duration $\tau_c$ time instants, as shown in Fig.~\ref{ResourceBlock}, is divided into three phases: i) $\tau_u$-instant uplink (UL) pilot transmission; ii) $\tau_d$-instant DL pilot transmission; and iii) $\tau_c-(\tau_u+\tau_d)$-instant data transmission. These phases are discussed below: 
\subsubsection{UL Training Interval} 
All UEs transmit UL pilots and the APs estimate UL channels of UEs within their~clusters.
\subsubsection{DL Training Interval}
The UEs herein, as discussed earlier,  estimates the effective DL CSI.
\vspace{-5pt}
\begin{figure}[h]
	\includegraphics[width=0.8\linewidth]{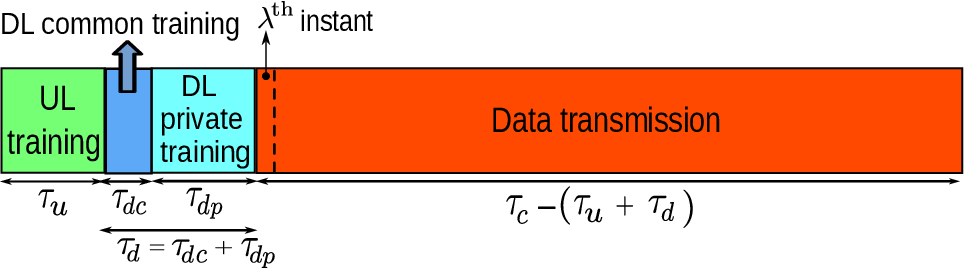}
	\vspace{-5pt}
	\caption{{Resource block schematics for DL training based CF RSMA system}.}
	\label{ResourceBlock}
\end{figure} 
To aid the UE DL CSI estimation, each AP,  as shown in Fig.~\ref{ResourceBlock}, transmits precoded pilots in two sub-phases: common training sub-phase 1 of $\tau_{dc}$ time instants, and private  training sub-phase 2 of $\tau_{dp}$ time instants.
\subsubsection{Data Transmission Interval} All APs transmit the precoded RSMA signal to the UEs in their clusters. 
Each UE decodes its common message by performing SIC of common messages from other clusters. {This SIC, as shown later in Fig.~\ref{fig3b_Vs_UEs}, suppresses MCI  in user-centric CF RSMA systems and} improves the system SE. Each UE then decodes its private signal by performing  SIC of its common message.
\vspace{-4pt}
\subsection{Channel Model With Aging}
The channel from the $k$th UE to the $m$th AP at the time instant $\lambda$ is denoted as $\hv_{mk}[\lambda] \in \mathbb{C}^{N\times 1}$. Due to densely deployed APs, the channel $\hv_{mk}[\lambda]$ includes both LoS and  non-LoS components. It is accordingly modeled as follows \cite{Zhang2024}:
\begin{align}
	\hv_{mk}[\lambda] = {\bar \hv}_{mk} e^{j\phi_{mk}^{\lambda}}+\Rmat_{mk}^{\frac{1}{2}} {\breve \hv}_{mk}[\lambda].\label{channel}
\end{align}
Here, ${\bar \hv}_{mk}=\sqrt{\frac{{K}_{mk}\beta_{mk}}{1+\mathcal{K}_{mk}}} \hv_{mk}^{LoS}$, and $\Rmat_{mk}={\frac{\beta_{mk}}{1+{K}_{mk}}} \widetilde \Rmat_{mk}$. The scalars ${K}_{mk}$ and $\beta_{mk}$ denote the Rician factor and the large-scale fading coefficient, respectively. The vectors $\hv_{mk}^{LoS}\in\mathbb{C}^{N\times 1}$ and ${\breve \hv}_{mk}[\lambda]\sim\mathcal{CN}(\mathbf{0},\Imat_N)$ models the LoS component and the small-scale NLoS component, respectively. The matrix $\widetilde\Rmat_{mk}$ models the spatial correlation. The LoS phase-shift $\phi_{mk}^{\lambda}$ is uniformly distributed over $[-\pi,\pi]$. The long-term channel statistics \ie $\bar \hv_{mk}$  and $\Rmat_{mk}$, remain constant over a resource block and, similar to~\cite{Zhang2024}, are assumed to be known at the AP. We assume, similar to~\cite{Zhang2024}, that $\phi_{mk}^{\lambda}$ varies as rapidly as the small-scale fading. Since the AP is unaware of the phase shift $\phi_{mk}^{\lambda}$, it estimates   $\hv_{mk}$ without assuming  $\phi_{mk}^{\lambda}$ knowledge. 
\subsubsection*{Channel Aging}
The channel in a resource block, due to UE mobility, evolves with time. The time-varying channel, at the time instant $t \in \{1, \cdots, \tau_c\}$ is modeled using Jake's model as a weighted sum of the channel at the time instant $\lambda$, and an independent innovation component, as follows~\cite{Chen2025}:
\begin{align}
	\hv_{mk}[t]=\rho_{t}\hv_{mk}[\lambda]+\bar \rho_{t} \bar \fv_{mk}[t].\label{channel_aging}
\end{align}   
The scalar $\bar \rho_{t}=\sqrt{(1-\rho_{t}^{2})}$, where $\rho_{t}=J_0(2\pi f_{d,i}T_s(t\lambda))$ is temporal correlation coefficient \cite{Chen2025}. Here, $J_0(\cdot)$ is the zeroth-order Bessel function, $T_s$ is the sampling time and $f_{d,i} = (v_kf_c)/c$ denotes the Doppler spread, with $v_k$, $c$ and $f_c$ being the speed of the UE and the light, and carrier frequency, respectively. The innovation component $\bar \fv_{mk}[t]$ is independent of ${\hv}_{mk}[\lambda]$, and also identically distributed as  ${\hv}_{mk}[\lambda]$.
\vspace{-8pt}
\subsection{UL training phase}
In  the UL training phase, the $k$th UE transmits the pilot $\sqrt{\tilde p_u} \phi_{k}[t_k]$ at the time instant $t_k$ for $1 \leq t_k \leq \tau_u$, where $\tilde p_u$ is the UL pilot power and $|\phi_{k}[t_k]|^2 = 1$. We assume that UEs within a cluster use orthogonal pilots. The UEs across cluster, however,  reuse pilots, which leads to pilot contamination~\cite{Zhang2024}. Let $\mathcal{P}_{k}$ be the set of UEs from different clusters which transmit pilot at the same instant $t_k$. The pilot symbol received by the $m$th AP from the $k$th UE at the time instant $t_k$ is  given as:
\begin{align}
	\yv_{m}[t_k]=\sum_{i\in\mathcal{P}_k}\hv_{mi}[t_k]\sqrt{\tilde p_u}\phi_{k}[t_k]+{\nv}_m[t_k]. \label{UL_training}
\end{align}
The vector ${\nv}_m[t_k]$, with probability density function (pdf) $\mathcal{CN}(\mathbf{0}, \sigma_m^2 \Imat_N)$, is the receiver noise at the $m$th AP. The received signal $\yv_{m}[t_k]$ can be used to estimate CSI at any time instant within the resource block. The CSI estimate quality, however, degrades as the time gap increases between the pilot transmission instant $(1<t_k<\tau_u+\tau_d)$ and CSI estimation instant $(\tau_u+\tau_d+1<n<\tau_c)$. Without loss of generality, we assume that each AP estimates UL channel of UEs in its cluster at the beginning of data transmission \ie $\lambda=\tau_u+\tau_d+1$.
The received signal in \eqref{UL_training} can be written using  \eqref{channel_aging},~as follows:
\begin{align}
	\yv_{m}[t_k]&=\sum_{i\in\mathcal{P}_k}(\rho_{t_k}\hv_{mi}[\lambda]+\bar \rho_{t_k}{\bar \fv}_{mi}[t_k])\sqrt{\tilde p_u}\phi_{k}[t_k]+\nv_m[t_k]. \nonumber
\end{align}
Using $\yv_{m}[t_k]$, the UL CSI estimate ${\hat \hv}_{mk}[\lambda]$ is given as \cite{Chen2025}:
\begin{align}
	{\hat \hv}_{mk}[\lambda]=\rho_{t_k}\sqrt{\tilde p_u}\overline \Rmat_{mk}\mathbf{\Psi}_{mk}\yv_{m}[t_k]. \label{UL_estimate_formula}
\end{align} 
The matrices ${\mathbf{\Psi}_{mk}}=\big(\sum_{i\in\mathcal{P}_k}\tilde p_u \overline \Rmat_{mi}+\sigma_{m}^2\Imat\big)^{-1}$, $\overline \Rmat_{mk}=\bar \hv_{mk}\bar \hv_{mk}^{H}+\Rmat_{mk}$.  The estimate
$\hat \hv_{mk}$ and estimation error $\tilde \hv_{mk}=\hv_{mk}-\hat \hv_{mk}$ are uncorrelated, and $\E[\hat{\hv}_{mk}\hat{\hv}_{mk}^H] = \widehat \Rmat_{mk} \triangleq \tilde p_{u}\rho_{t_k}^2\overline \Rmat_{mk}\mathbf{\Psi}_{mk}^{-1}\overline \Rmat_{mk}$. 
\vspace{-5pt}
\subsection{{DL Training}} 
We now explain two stages of DL training as follows.
\subsubsection{DL common training} Since each cluster has one common message, the APs in each cluster send a single precoded common pilot to their respective UEs. The DL common training sub-phases, consequently,  has $\tau_{dc}=L$ time instants, corresponding to the $L$ clusters. {The $L$ orthogonal pilots across clusters in this phase  avoid pilot contamination.} In the $\tilde t_l$th time instant, the $m$th AP in the set $\mathcal{A}_l$ transmits the precoded pilot symbol $\sqrt{ p_{lm}^{c}}\vv_{lm}^{c}\phi_{c}[\tilde t_l]$. The scalar $\phi_{c}[\tilde t_l]$ is the pilot symbol for the $l$th cluster, with $|\phi_{c}[\tilde t_l]|^2=1$ The scalar $p_{lm}^{c}$ is the DL common pilot power. The vector {$\vv_{lm}^{c}$} is the normalized common precoder for the $m$th AP.  
The DL common pilot signal received at the $k$th UE in the $l$th cluster from all the APs in the set $\mathcal{A}_l$ at the time instant $\tilde t_l$, is given as follows:
\begin{align}
	y_{lk}^{c}[\tilde t_l]&=\sum_{m\in\mathcal{A}_l}\hv^{H}_{mk}[\tilde t_l]\sqrt{p_{lm}^{c}}\vv_{lm}^{c}\phi_{c}[\tilde t_l]+n_{k}[\tilde t_l] \nonumber \\
	&\triangleq a_{lk}^{c}[\tilde t_l]\phi_{c}[\tilde t_l]+n_{k}[\tilde t_l].\label{DL_pilot_common}
\end{align} 
The scalar $a_{lk}^{c}[\tilde t_l]$ denotes the effective DL common channel of the $k$th UE in the $l$th cluster at the $\tilde t_l$th instant. The scalar $n_{k}[\tilde t_l]$, with pdf $\mathcal{CN}(0,\sigma_k^2)$, is the receiver noise at the $k$th UE. Using \eqref{channel_aging}, the scalar $a_{lk}^{c}[\tilde t_l]$ is expressed as follows: $a_{lk}^{c}[\tilde t_l]=$
\begin{align}
	&\rho_{\tilde t_l}\sum_{m\in\mathcal{A}_l}\sqrt{p_{lm}^{c}} \hv_{mk}^H[\lambda] \vv_{lm}^{c}+\bar \rho_{\tilde t_l}\sum_{m\in\mathcal{A}_l}\sqrt{p_{lm}^{c}} \bar \fv_{mk}^H[\tilde t_l]\vv_{lm}^{c}\nonumber \\
	&\triangleq \rho_{\tilde t_l}a_{lk}^{c}[\lambda]+\bar\rho_{\tilde t_l} z_{lk}^{c}[\tilde t_l].\label{DL_agingchannel_common}
\end{align} 
The term $z_{lk}^{c}$ 
represents the $\tilde t_l$th instant innovation component in the effective DL common channel due to channel aging. 
The scalar $y_{lk}^{c}[\tilde t_l]$ in \eqref{DL_pilot_common} is the sufficient statistic to estimate the DL common CSI $ a_{lk}^{c}[\lambda]$.
We next derive the LMMSE estimate of $a_{lk}^{c}[\lambda]$ in the proposition, which is proved in~Appendix~\ref{DL common channel estimation}. 

\begin{Proposition} \label{proposition_common_DL_estimate}
	The LMMSE estimate of the DL common channel $a_{lk}^{c}[\lambda]$ in a clustered user-centric CF system with spatially-correlated Rician fading channels is given as follows
	\begin{align}
		\hat a_{lk}^{c}[\lambda]&=\bar a_{lk}^{c}[\lambda]+\theta_{lk}^{c}(\psi_{lk}^{c})^{-1}\big(y_{lk}^{c}[\tilde t_l]-\bar y_{lk}^{c}[\tilde t_l]\big).  \label{DL_common_estimate_forluma}
	\end{align}
	The simplified closed-form expressions for the scalars $\bar a_{lk}^{c}[\lambda]$, $\theta_{lk}^{c}$, $\psi_{lk}^{c}$ and $\bar y_{lk}^{c}[\tilde t_l]$ are given in Appendix~\ref{DL common channel estimation}. The covariances of estimated DL common CSI $\hat a_{lk}^{c}[\lambda]$ and the estimation error $\tilde a_{lk}^{c}[\lambda]=a_{lk}^{c}[\lambda]-\hat a_{lk}^{c}[\lambda]$ are given as $\E\{|\hat a_{lk}^{c}[\lambda]-\bar a_{lk}^{c}[\lambda] |^2\}\triangleq \hat r_{lk}^{c}=\theta_{lk}^{c}(\psi_{lk}^{c})^{-1}\theta_{lk}^{c}$ and $\E\{|\tilde a_{lk}^{c}[\lambda]|^2\}=\tilde r_{lk}^{c}=r_{lk}^{c}-\hat r_{lk}^{c}$, whose simplified expressions are derived in Appendix~\ref{DL common channel estimation}.
\end{Proposition}
Each UE can similarly estimate the effective DL common channels from APs in other clusters \ie $\{\hat{a}_{l'k}^{c}\}$ by correlating the receive pilot signal in \eqref{DL_pilot_common} with respective pilot sequence $\phi_{c}[\tilde{t}_{l'}]$. This information is later used to suppress MCI from other clusters while decoding the common data.
\subsubsection{DL private training}
In this phase, all the APs transmits the pilot symbols to the $k$th UE in their corresponding cluster at the time instant $\tilde t_k=\tau_u+\tau_{dc}+t_k$, such that $\tilde t_k\in \tau_u+[\tau_{dc}\; \tau_{d}]$. We assume that both UL and DL pilots for the $k$th UE are same \ie $\phi_{k}[\tilde t_k]=\phi_{k}[t_k]$. The APs transmit  pilots at different time instants. For example, the $m$th AP in $\mathcal{A}_l$ transmits the pilot $\sqrt{ p_{mk}^{p}}\vv_{mk}^{p}\phi_{k}[\tilde t_k]$ to the $k$th UE in $\mathcal{K}_l$ at the time instant $\tilde t_k$. The scalars $p_{mk}^{p}$ is  DL private pilot power, and $\vv_{mk}^{p}$ is the normalized private precoder.
The DL private pilot signal received by the $k$th UE in the $l$th cluster from all APs at time instant $\tilde t_k$ is, accordingly, given as follows:
\begin{align}
	&y_{lk}^{p}[\tilde t_k]=\sum_{m\in \mathcal{A}_l}\hv^{H}_{mk}[\tilde t_k]\sqrt{p_{mk}^{p}} \vv_{mk}^{p}\phi_{k}[\tilde t_k]\nonumber \\
	&+\sum_{l'\neq l}^{L}\sum_{n\in \mathcal{A}_{l'}}\hv^{H}_{nk}[\tilde t_k]\sum_{i\in\mathcal{P}_k}\sqrt{ p_{ni}^{p}}\vv_{ni}^{p}\phi_{k}[\tilde t_k]+ n_{k}[\tilde t_k] \nonumber \\
	&\triangleq a_{lkk}^{p}[\tilde t_k]\phi_{k}[\tilde t_k]+\sum_{l'\neq l}^{L}\sum_{i\in\mathcal{P}_k}a_{l'ki}^{p}[\tilde t_k]\phi_{k}[\tilde t_k]+n_{k}[\tilde t_k].\label{DL_pilot_private} 
\end{align} 
The scalar $a_{lki}^{p}[\tilde t_k]$ is the effective DL private channel of the $k$th UE in the $l$th cluster at the $\tilde t_k$th instant. Similar to \eqref{DL_agingchannel_common}, the effective DL private channel is: $	a_{lki}^{p}[\tilde t_k]=$
\begin{align}
	&\rho_{\tilde t_k}\sum_{m\in\mathcal{A}_l}\sqrt{p_{mi}^{p}}\hv_{mk}^H[\lambda]\vv_{mi}^{p}
	+\bar \rho_{\tilde t_k}\sum_{m\in\mathcal{A}_l}\sqrt{p_{mi}^{p}}\bar \fv^{H}_{mk}[\tilde t_k]\vv_{mi}^{p}\nonumber \\
	&\triangleq\rho_{\tilde t_k}a_{lki}^{p}[\lambda]+\bar \rho_{\tilde t_k}z_{lki}^{p}[\tilde t_k].\label{DL_agingchannel_private} 
\end{align}
The scalar $z_{lki}^{p}[\tilde t_k]$ 
represents the $\tilde t_k$th instant innovation component in the effective DL private channel, due to channel aging.
The scalar $y_{lk}^{p}[\tilde t_k]$ in \eqref{DL_pilot_private} is the sufficient statistic to estimate the DL private CSI $\hat a_{lki}^{p}[\lambda]$, which we derive in the following proposition, which is proved in Appendix~\ref{DL private channel estimation}. 
\begin{Proposition} \label{proposition_private_DL_estimate}
	The LMMSE estimate of the DL private channel $a_{lki}^{p}[\lambda]$ in a clustered CF system with spatially-correlated Rician fading channels is given as follows:
	\begin{align}
		\hat a_{lki}^{p}[\lambda]&=\bar a_{lki}^{p}[\lambda]+\theta_{lki}^{p}(\psi_{lki}^{p})^{-1}\big(y_{lk}^{p}[\tilde t_k]-\bar y_{lk}^{p}[\tilde t_k]\big). \label{DL_private_estimate_forluma}
	\end{align}
	The closed form expressions of $\bar a_{lki}^{p}[\lambda]$, $\theta_{lki}^{p}$, $\psi_{lki}^{p}$ and $\bar y_{lk}^{p}[\tilde t_k]$ are given in Appendix~\ref{DL private channel estimation}. The estimated DL private CSI $\hat a_{lki}^{p}[\lambda]$ and the estimation error $\tilde a_{lki}^{p}[\lambda]=a_{lki}^{p}[\lambda]-\hat a_{lki}^{p}[\lambda]$ are uncorrelated \cite{Yao2024}. 
	We now discuss data transmission.
	
\end{Proposition} 
%
%
%
%
%
%
%
%
%
%
%
%
\subsection{RSMA Data Transmission}
Let, $x_{l}^{c}[t]$ be the common data signal for the $l$th cluster, and let $x_k[t]$ be the private data signal of the $k$th UE at time instant $t$. The APs in the $l$th cluster transmit the precoded common signal $\sqrt{p_{lm}^{c}}\vv_{lm}^{c}x_{l}^{c}[t]$ to UEs in $\mathcal{K}_l$ and precoded private data signal $\sqrt{p_{mk}^{p}}\vv_{mk}^{p}x_k[t]$ to the $k$th UE in $\mathcal{K}_l$. The scalars $p_{lm}^{c}$ and $p_{mk}^{p}$ are transmit powers of common and private data symbols respectively, with $|x_{l}^{c}[t]|^2=|x_{k}[t]|^2=1$.  The vectors $\vv_{lm}^{c}=\sqrt{\eta_{lm}^{c}}\sum_{i\in\mathcal{K}_l}\hat \hv_{mi}[\lambda]$ and $\vv_{mk}^{p}=\sqrt{\eta_{mk}^{p}}\hat \hv_{mk}[\lambda]$ are common and private precoders respectively, with $\eta_{lm}^{c}=\frac{1}{\sqrt{\E[\|\vv_{lm}^{c}\|^2]}}$ and $\eta_{mk}^{p}=\frac{1}{\sqrt{\E[\|\vv_{mk}^{p}\|^2]}}$ being the normalization factors. In the DL data transmission phase, each UE receives signals from all clusters. The signal received by the $k$th UE in the $l$th cluster at the $t$th time instant~is given as :   
\begin{align}
	\hspace{0pt}&y_{lk}^{}[t]=\underbrace{\sum_{l'=1}^{L}\sum_{m\in\mathcal{A}_{l'}}\sqrt{p_{l'm}^{c}}\,\hv_{mk}^H[t]\vv_{l'm}^{c}[\lambda]x_{c,{l'}}[t]}_{\text{common signal from all clusters}}+\underbrace{n_k[t]}_{\text{receiver noise}}\nonumber
	\end{align}
	\begin{align}
	\hspace{0pt}&+\underbrace{\sum_{l'=1}^{L}\sum_{m\in\mathcal{A}_{l'}}\sum_{j\in\mathcal{K}_{l'}}\sqrt{p_{mj}^{p}}\,\hv_{mk}^H[t]\vv_{mj}^{p}[\lambda]x_{j}[t]}_{\text{private signal of all UEs}}.\label{UE_receive_signal}
\end{align}

Using the definitions of effective DL common and private channels $a_{lk}^{c}[t]$ and $a_{lkj}^{p}[t]$ in \eqref{DL_pilot_common} and \eqref{DL_pilot_private} respectively, the UE receive signal in \eqref{UE_receive_signal} can be simplified as follows:
\begin{align}
	\hspace{-3pt}y_{lk}^{}[t]&=\sum_{l=1}^{L}a_{lk}^{c}[t]x_{l}^{c}[t] +\sum_{l'=1}^{L}\sum_{j\in\mathcal{K}_{l'}}a_{l'kj}^{p}[t]x_{j}[t]+n_k[t].
	\label{UE_received_Signal}
\end{align}
\vspace{-10pt}
\subsection{RSMA Decoding at the UE} 
\vspace{-5pt}
\subsubsection{Common Signal Decoding} To decode its common signal, each UE performs SIC to remove the MCI due to common data by using the effective DL channel estimates $\hat a_{l'k}^{c}[\lambda]\,\forall l'\in\{1,\cdots,L\}, l'\neq l$. Due to CSI estimation error,  the SIC of MCI is imperfect and residual MCI remains. The UE receive signal in \eqref{UE_received_Signal}, after SIC, is written to decode its common symbol $x_l^c[t]$ as follows:
\begin{align}
	{y}_{lk}^{c}[t]&= \underbrace{a_{lk}^{c}[t]x_{l}^{c}[t]}_{\substack{\text{common signal from}\\\text{own cluster}}}+\underbrace{\sum_{l'=1,l'\neq l}^{L}(a_{l'k}^{c}[t]-\hat a_{l'k}^{c}[\lambda])x_{l'}^{c}[t]}_{\text{residual interference}}\nonumber\\
	&+\underbrace{\sum_{l'=1}^{L}\sum_{j\in\mathcal{K}_{l'}}a_{l'kj}^{p}[t]x_{j}[t]}_{\text{private signal of all UEs}}+\underbrace{n_k[t]}_{\text{receiver noise}}\hspace{-5pt}.\label{common_signal_desired}	
\end{align}
After expanding $a_{lk}^{c}[t]$ by using \eqref{DL_agingchannel_common}, and by substituting $a_{lk}^{c}[\lambda]=\hat a_{lk}^{c}[\lambda] +\tilde a_{lk}^{c}[\lambda]$, the receive signal $y_{lk}^{c}[t]$ can be expressed as follows:
\begin{align}
	y_{lk}^{c}[t]
	&=\underbrace{\rho_{t}\hat a_{lk}^{c}[\lambda]x_{l}^{c}[t]}_{\text{desired common signal, }N_{lk,t}^{c}} + \underbrace{n_{lk}^{c}[t]}_{\text{effective common noise}}. \label{common_signal_desired_AfterSIC}
\end{align}
The effective common noise $n_{lk}^{c}[t]$ is given in \eqref{common_signal_effective_noise}.\subsubsection{Private Data}
The $k$th UE, to decode its private data, performs SIC to cancel the common data of the $l$th cluster from the receive signal in \eqref{common_signal_desired}. The UE received signal is
\begin{figure*}{}
	{ \begin{align} 
			n_{lk}^{c}[t]&=\underbrace{\rho_{t}\tilde a_{lk}^{c}[\lambda]x_{l}^{c}[t]}_{\substack{\text{common channel}\\ \text{estimation error, } D_{lk_1,t}^{c}} }+\underbrace{\bar \rho_{t} z_{lk}^{c}[ t]x_{l}^{c}[t]}_{\substack{\text{common channel}\\ \text{ aging,} D_{lk_2,t}^{c} }}+\underbrace{\sum_{l'=1,l'\neq l}^{L}(a_{l'k}^{c}[t]-\hat a_{l'k}^{c}[\lambda])x_{l'}^{c}[t]}_{\text{residual interference, }D_{lk_3,t}^{c} }+\underbrace{\sum_{l'=1}^{L}\sum_{j\in\mathcal{K}_{l'}}a_{l'kj}^{p}[t]x_{j}[t]}_{\text{private signal from all cluster, }D_{lk_4,t}^{c}}+\underbrace{n_k[t]}_{\substack{\text{receiver noise, }\\D_{lk_5,t}^{c}}}\hspace{-10pt}.\label{common_signal_effective_noise}
	\end{align} }\hrule \vspace{-5pt}
\end{figure*}  
\begin{align}
	y_{lk}^{p}[t]
	&=\underbrace{a_{lkk}^{p}[t]x_{k}[t]}_{\text{private signal}}  +\underbrace{\sum_{l'=1}^{L}\sum_{j\in\mathcal{K}_{l'},j\neq k}a_{l'kj}^{p}[t]x_{j}[t]}_{\text{private interference}}\nonumber\\
	&+ \underbrace{\sum_{l'=1,l'\neq l}^{L}(a_{l'k}^{c}[t]-\hat a_{l'k}^{c}[\lambda])x_{l'}^{c}[t]}_{\text{residual interference}} +\underbrace{n_k[t]}_{\text{receiver noise}}\hspace{-10pt}.
\end{align}
The UE receive signal, after expanding $a_{lkk}^{p}[t]$ using \eqref{DL_agingchannel_private} and by substituting $a_{lkk}^{p}[\lambda]=\hat a_{lkk}^{p}[\lambda]+\tilde a_{lkk}^{p}[\lambda] $ in the first term, can be expressed as follows:
\begin{align}
	{y_{lk}^{p}[t]}&=\underbrace{\rho_t\hat a_{lkk}^{p}[\lambda]x_{k}[t]}_{\text{desired private signal, }N_{lk}^{p}[t]}+\underbrace{n_{lk}^{p}[t]}_{\text{effective private noise}}\hspace{-20pt}. \label{private_signal_desired}
\end{align}
The expression of effective private noise $n_{lk}^{p}[t]$ is given in \eqref{private_signal_effective_noise}. 
We now provide the SE expression for  CF mMIMO RSMA system with channel aging.
\begin{figure*}{}
	{\begin{align} 
			n_{lk}^{p}[t]=\underbrace{\rho_t\tilde a_{lkk}^{p}[\lambda]x_{k}[t]}_{\substack{\text{private channel}\\ \text{estimation error, } D_{lk_1,t}^{p}} }+\underbrace{\bar \rho_tz_{lkk}^{p}[t]x_{k}[t]}_{ \substack{\text{private channel} \\\text{aging, } D_{lk_2,t}^{p}} } +\underbrace{\sum_{l'=1}^{L}\sum_{j\in\mathcal{K}_{l'}, j\neq k}a_{l'kj}^{p}[t]x_{j}[t]}_{\text{private interference, }D_{lk_3,t}^{p}} + \underbrace{\sum_{l'=1,l'\neq l}^{L} (a_{l'k}^{c}[t]-\hat a_{l'k}^{c}[\lambda])x_{l'}^{c}[t]}_{\text{residual interference, }D_{lk_4,t}^{p}} +\underbrace{n_k[t]}_{\substack{\text{receiver noise, }\\D_{lk_5,t}^{p}}}\hspace{-10pt}. \label{private_signal_effective_noise}
	\end{align}}\hrule \vspace{-5pt}
\end{figure*}  
\vspace{-5pt}
\subsection{Achievable Spectral Efficiency}
The sum SE of the CF mMIMO RSMA system by considering channel aging is given as 
\begin{align}
	\text{SE}_{}^{erg}&=\frac{1}{\tau_c}\sum_{t=\lambda}^{\tau_c}\Big(\sum_{l=1}^{L}\text{SE}_{l}^{c}[t]+\sum_{k=1}^{K}\text{SE}_{lk}^{p}[t] \Big), \text{where} \label{SE_erg}	\\
	\text{SE}_{l}^{c}[t]&=\underset{\substack{k\in \mathcal{K}_l}}{\text{ min}}\text{SE}_{lk}^{c}[t] =\underset{\substack{k\in \mathcal{K}_l}}{\text{ min}}\E\{\log_2(1+\Gamma_{lk,t}^{c})\} \text{ and } \label{Common_SimulatedSE} \\
	\text{SE}_{lk}^{p}[t]&=\E\{\log_2(1+\Gamma_{lk,t}^{p})\}. \label{Private_SimulatedSE}
\end{align}
The term $\text{SE}_{l}^{c}[t]$ denotes  the common SE of the $l$th cluster, and  $\text{SE}_{lk}^{p}[t]$ is the private SE of the $k$th UE in $\mathcal{K}_l$ at the $t$th time instant. The scalars $\Gamma_{lk,t}^{c}$ and $\Gamma_{lk,t}^{p}$ denote the common and private SINR respectively, of  the $k$th UE at the time instant $t$. Their expressions, by using \eqref{common_signal_desired_AfterSIC} and \eqref{private_signal_desired}, are given as follows: 
\begin{align}
	&\Gamma_{lk,t}^{c}= \frac{|N_{lk,t}^{c}|^2}{\sum_{i=1}^{5}|D_{lk_i,t}^{c}|^2} \text{ and}  &\Gamma_{lk,t}^{p}= \frac{|N_{lk,t}^{p}|^2}{\sum_{i=1}^{5}|D_{lk_i,t}^{p}|^2}. \label{SINR_definitions}	
\end{align}
{We, therefore, numerically simulate these SE expressions to understand the necessity of downlink training in user-centric CF mMIMO RSMA systems, which we do in the next section.}
\section{Simulation Results}
We now investigate the performance of the RSMA system
by comparing it with its SDMA counterpart in presence of
channel aging, pilot contamination and spatially-correlated
Rician channels. For these studies, we consider a CF system with a coverage area of $1\times1$ Km$^2$. All APs and UEs are randomly located, and are clustered into $L$ clusters based on {the DCC protocol~\cite{demir2021}}. The coverage area is wrapped around the edges to eliminate boundary effects. We consider a bandwidth of $B = 20$~MHz, and a resource block of length $\tau_c=100$. The correlation matrix $\Rmat_{mk}$ is modeled using the Gaussian local scattering model, with an ASD of $30^\circ$~\cite{demir2021}. The large scale fading coefficient $\beta_{mk}$ and the Rician factor ${K}_{mk}$ for the UE-to-AP channels are modeled as in~\cite[Table-5.1]{3gpp2025}. We split,  similar to \cite{Flores2023},  common and private powers as $p_{lm}^{c}=(1-t_m)p^{max}$ and $p_{mk}^{p}=t_{m}\frac{p^{max}}{\widetilde K_l}$, with a power splitting factor of $t_m=0.05$. The noise variance $\sigma_m^2=\sigma_k^2=-94$~dBm. We set $N=4$ AP antennas, $M=16$ APs and UL pilot power $\tilde p_{u}=30$~dBm. All these parameters remain fixed unless explicitly specified. 
\begin{figure*}[htbp]
	\begin{subfigure}{0.28\textwidth}
		\includegraphics[width = \linewidth, height=\linewidth]{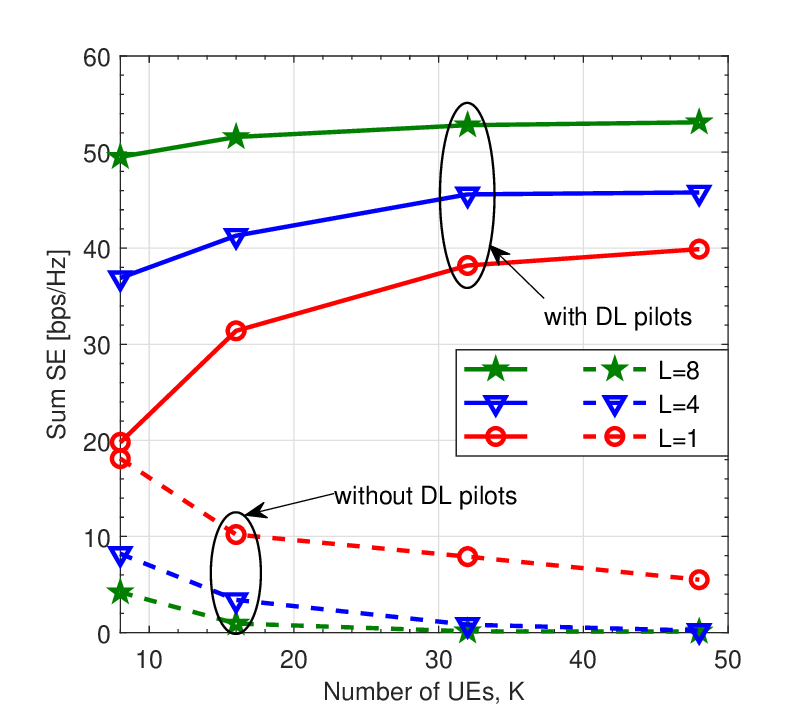}
		\caption{}
		\label{fig3b_Vs_UEs}
	\end{subfigure}
	\begin{subfigure}{0.28\textwidth}\vspace{-15pt}
		\includegraphics[width = \linewidth, height=\linewidth]{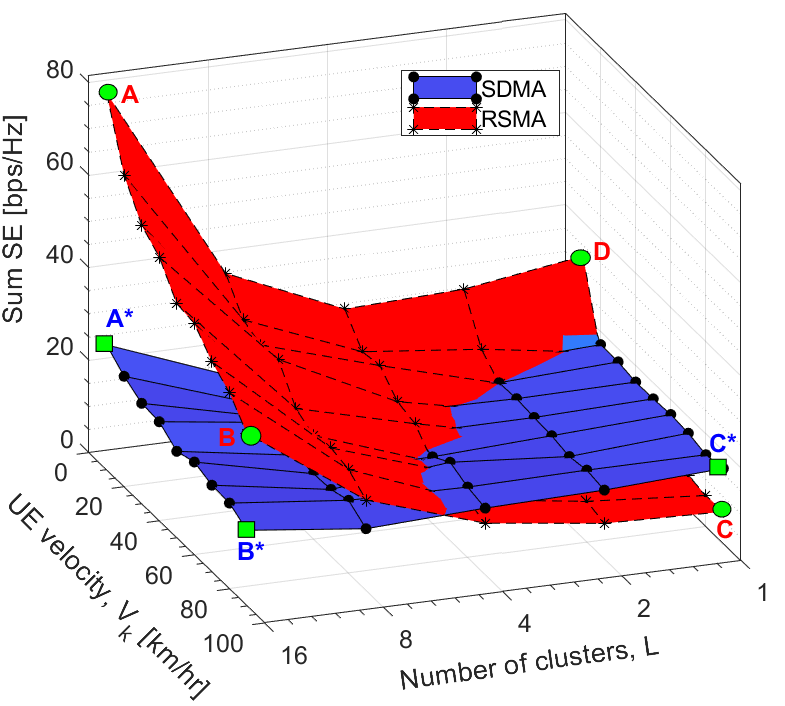}\vspace{-5pt}
		\caption{}
		\label{fig3c_Vs_Vel_Clusters}
	\end{subfigure}
	\begin{subfigure}{0.28\textwidth}\vspace{-4pt}
		\includegraphics[width = \linewidth, height=\linewidth]{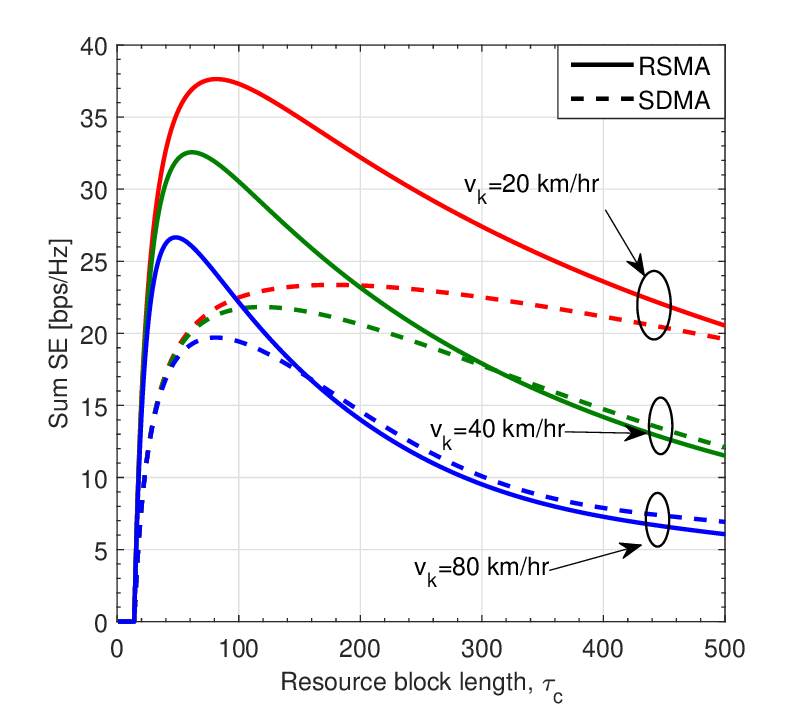}\vspace{-7pt}
		\caption{}
		\label{fig3d_Vs_tau_c}
	\end{subfigure}
	\vspace{-0.35cm}
	\caption{RSMA SE, with and without DL pilots, by varying the number of UEs $K$; RSMA and SDMA comparison by varying    
		b) number of clusters $L$ and UE velocity {$v_k$}; and c) resource block length $\tau_c$ for different UE velocity $v_k$. 	
	}
\end{figure*}
\subsubsection{Downlink Pilots and Clustering to Handle Higher Number of UEs}
We now compare in Fig.~\ref{fig3b_Vs_UEs}, the RSMA sum SE with and without DL pilots by varying the total number of UEs $K$ in the system, for $L=1,4$ and $8$ clusters. For RSMA without DL pilots, each UE employs channel statistics $\bar a_{lk}^{c}[\lambda]$ for performing SIC of the common data instead of the estimate $\hat a_{lk}^{c}[\lambda]$ ~\cite{demir2021}, and has $\tau_c-\tau_u$ time instants instead of $\tau_c-\tau_u-\tau_d$ for data transmission. We fix $M=32$ APs, UE velocity $v_k=40$ km/hr, and AP transmit power $p^{\text{max}}=30$ dBm. 
We first see that the sum SE with DL pilots increases with  increased number of clusters $L$. This is because the common sum SE, defined in \eqref{Common_SimulatedSE}, is the minimum SE of all UEs in a cluster, which is summed over all clusters. As $L$ increases, the common SE per cluster improves due to improved SIC, and reduced interference from private messages. This increases the sum SE. The sum SE without DL pilots, however, decreases with $L$. This is due to the increased MCI, which increases with increased $L$ value. Without DL pilots, the UEs rely on the statistical CSI $\bar{a}_{lk}^c[\lambda]$, which limits their ability to perform efficient SIC of the common data. \textit{We see that RSMA in user-centric CF mMIMO systems performs effectively only with DL pilots.}


\vspace{-4pt}
\subsubsection{{Joint Effect of Number of Clusters and UE velocity}} 
We compare in Fig.~\ref{fig3c_Vs_Vel_Clusters}, the SDMA and RSMA SE by varying the UE velocity, and the number of clusters $L$. The SE of SDMA is evaluated by setting $t_m=1$ ($p_{lm}^{c}=0$). We fix $K=16$ UEs, AP transmit power $p^{max}=30$~dBm, and vary $L$ from $1$ to $16$ clusters. We see that with increase in $L$, the RSMA SE increases rapidly than SDMA, even for high UE velocity (see point \text{C} to \text{B} and \text{D} to \text{A}). This is because the common sum SE increases with clusters $L$, which increases the RSMA sum SE as explained in Fig~\ref{fig3b_Vs_UEs}. We also see that for a UE velocity of $v_k = 100$~km/hr, RSMA outperforms SDMA for $L\geq4$, whereas opposite happens for $L<4$. This shows that the SE loss due to increased channel aging can be compensated by forming a higher number of clusters $L$. \textit{This study helps in determining when to prefer RSMA over SDMA, and the optimal UE clusters required for RSMA to outperform SDMA.} 
\subsubsection{Impact of Aging on Resource Block Length}
We now plot in Fig.~\ref{fig3d_Vs_tau_c}, the SE versus the resource block length $\tau_c$, and compare the SDMA and RSMA SE for different UE velocities. We fix $L=8$ clusters, and $K=24$ UEs. {We see that for different velocities $v_k$, RSMA outperform SDMA for different $\tau_c$ values. And this $\tau_c$ value for which RSMA outperforms SDMA, reduces with increase in $v_k$. These happens because RSMA performance critically depends on the SIC. The SIC quality, in turn,  depends on the DL CSI estimate, whose quality degrades with increase  in $v_k$, and consequently $\tau_c$.  Further, the marginally lower SE of RSMA than SDMA for high $\tau_c$ values is due to the higher estimation overhead (common DL training phase) in RSMA than SDMA. \textit{This analysis, thus,  helps to choose the right $\tau_c$ values required for RSMA to outperform SDMA for a given UE velocity.}} 

\vspace{-2pt}
\section{Conclusion}
We evaluated  the ergodic SE of the RSMA-based user-centric CF mMIMO system with
spatially-correlated Rician fading channels with aging. We showed that the  DL CSI estimation is critical for a user-centric CF mMIMO system to realize the RSMA SE gains. We showed that channel aging conspicuously degrades the RSMA gain. We also showed that by increasing the number of clusters, the RSMA gain over SDMA can be increased by compensating the degradation due to channel aging. We also investigated the impact of resource block length on RSMA, an analysis which can help in deciding the resource block length to maximize the SE gains.
\vspace{-3pt}
\appendices

\section{} \label{DL common channel estimation}
\subsection{DL Common Channel Estimation}
The LMMSE estimate of the effective DL common channel $a_{lk}^{c}[\lambda]$ is given as follows \cite{Yao2024}:
\begin{align}
	\hat a_{lk}^{c}[\lambda]&=\bar a_{lk}^{c}[\lambda]+\theta_{lk}^{c}(\psi_{lk}^{c})^{-1}\big(y_{lk}^{c}[\tilde t_l]-\bar y_{lk}^{c}[\tilde t_l]\big), \text{ where} \notag  
\end{align} 
 $\bar a_{lk}^{c}[\lambda]=\E\{a_{lk}^{c}[\lambda]\}=\sum\limits_{m\in\mathcal{A}_l}\sum\limits_{i\in \mathcal{K}_l}\sqrt{p_{lm}^{c}\eta_{lm}^{c}}\Tr(\Lambdamat_{mki}\widehat \Rmat_{mk})$,   
with $\Lambdamat_{mki}=\overline \Rmat_{mi}\overline \Rmat_{mk}^{-1}$  for $i\in \mathcal{P}_k$ and $0$ otherwise. The scalar $\bar y_{lk}^{c}[\tilde t_l]=\E\{y_{lk}^{c}[\tilde t_l]\}=\rho_{\tilde t_l}\bar a_{lk}^{c}[\lambda]$.
The covariance terms
$\theta_{lk}^{c}=\E\{(y_{lk}^{c}[\tilde t_k]-\bar y_{lk}^{c}[\tilde t_k])(a_{lk}^{c}-\bar a_{lk}^{c})^*\}$ and  $\psi_{lk}^{c}=\E\{|y_{lk}^{c}[\tilde t_k]-\bar y_{k}^{c}[\tilde t_k]|^2\} $ are calculated in \cite{RSMA_Tech_report}. Their simplified closed-form expressions are given in~Table~\ref{pi_terms}.
\begin{table*}[t]
	\footnotesize
	\caption{Simplified expressions for different terms present in $\hat a_{lk}^c[\lambda]$ and $\hat a_{lki}^p[\lambda]$.  }
	\centering
	\begin{tabular}{|c|}
		\hline
		\textbf{Terms in} $\theta_{lk}^c$, $\psi_{lk}^{c}$, $\theta_{lki}^p$ and $\psi_{lki}^{p}$ \\
		\hline 
		{$ \theta_{lk}^{c}=\sum_{m',m\in \mathcal{A}_{l}}\sum_{j,i\in \mathcal{K}_{l}}\sqrt{p_{lm'}^{c}p_{lm}^{c}}{\overline\eta_{lm'm}^{c}} \rho_{\tilde t_l}\Big({\pi_{m'mkji}- \varsigma_{m'mkji}}\Big)$} 
		\\
		
		\hline 
		
		{$\psi_{lk}^{c}=\sum_{m,m'\in\mathcal{A}_l}\sum_{j,i\in \mathcal{K}_l}\sqrt{p_{lm}^{c}p_{lm'}^{c}}{\overline\eta_{lmm'}^{c}}\Big( \rho_{\tilde t_l}^2(\pi_{m'mkji}-\varsigma_{m'mkji})
			+\bar \rho_{\tilde t_l}^2\alpha_{m'mkji} \Big) +\sigma_k^2 $ } \\ [0pt]
		\hline
		$\theta_{lki}^{p}=	\sum_{l'=1}^{L}\sum_{j\in\mathcal{P}_k}\sum_{ {m,m'\in\mathcal{A}_{l}}} \sqrt{p_{m'j}^{p}p_{mi}^{p}}{\overline \eta_{m'mji}^{p}}\rho_{\tilde t_k}\Big(\pi_{m'mkji}-\varsigma_{m'mkji} \Big)$
		\\[0pt]
		\hline 
		{$\psi_{lki}^{p}=\sum_{l,l'=1}^{L}  \sum_{j,i\in\mathcal{P}_k} \sum_{\substack{{m'\in\mathcal{A}_{l'}}\\ {m\in\mathcal{A}_{l}}}} \rho_{\tilde t_k}^{2} \sqrt{p_{m'j}^{p}p_{mi}^{p}}{\overline \eta_{m'mji}^{p}}\Big(\rho_{\tilde t_k}^{2}\pi_{m'mkji}+\bar \rho_{\tilde t_k}^{2}\alpha_{mkji}-\rho_{\tilde t_k}^{2}\varsigma_{m'mkji}\Big)+\sigma_k^2$}\\[4pt]
		\hline
		\hline
		{\textbf{Terms in $\pi_{m'mkji}$, $\varsigma_{m'mkji}$ and $\alpha_{m'mkji}$  }} \\ [1ex] 
		\hline 
		{$\pi_{m'mkji}= \begin{cases}
				\Tr(\Lambdamat_{m'kj}\widehat \Rmat_{m'k})\Tr(\Lambdamat_{mki}\widehat \Rmat_{mk}), &\text{if } m'\neq m  \\
				{\tilde p_u}\sum_{p\in\mathcal{P}_k}\Big(\rho_{t_k}^2{\pi_{mkjip}^{(1,1)}}+\bar \rho_{t_k}^2\pi_{mkjip}^{(2,1)}\Big)+\pi_{mkji}^{(3)}  , &\text{if } m'=m 
			\end{cases}$   } \vline \,\, {$\pi_{mkji}^{(3)}=\sigma_{m}^2\Tr(\Deltamat_{mj}\Deltamat_{mi}^H\overline \Rmat_{mk})$ }
		\\[4pt]
		\hline 
		{$\pi_{mkjip}^{(1,1)}= \begin{cases}
				\Tr(\Deltamat_{mj}\overline \Rmat_{mp}\Deltamat_{mi}^H\overline \Rmat_{mk}), &\text{if } p\neq k  \\
				\Tr(\overline \Rmat_{mk}\Deltamat_{mj}\overline \Rmat_{mk}\Deltamat_{mi}^H) +\bar \hv_{mk}^H\Deltamat_{mj}\bar \hv_{mk}\Tr(\Deltamat_{mi}^H \Rmat_{mk})+ \Tr(\Deltamat_{mj} \Rmat_{mk} ) \Tr(\Deltamat_{mi}^H \overline\Rmat_{mk} )  , &\text{if } p=k 
			\end{cases}$   }
		
		\\[4pt] 
		
		\hline
		$\pi_{mkjip}^{(2,1)}= \begin{cases}
			\Tr\big(\Rmat_{mk}\Deltamat_{mj}\Rmat_{mp}\Deltamat_{mi}^H \big)
			, &\text{if  } p\neq k  \\
			\bar \hv_{mk}^H\Deltamat_{mj}\bar \hv_{mk}\bar \hv_{mk}^H\Deltamat_{mi}^H\bar \hv_{mk}+\Tr\big(\Rmat_{mk}\Deltamat_{mj}\Rmat_{mk}\Deltamat_{mj}^H \big) +\Tr\big(\Deltamat_{mj}\Rmat_{mk} \big) \Tr\big( \Deltamat_{mi}^H\Rmat_{mk}\big), &\text{if  } p=k.
		\end{cases}$  \\[4pt]

		\hline
		{$ \varsigma_{m'mkji} ={ \Tr(\Lambdamat_{mki}\widehat \Rmat_{mk}) \Tr(\Lambdamat_{m'kj}\widehat \Rmat_{m'k})  }$}	\vline\,\,{$\alpha_{m'mkji}=\begin{cases}
				0, &\text{if } m'\neq m  \\
				\sum_{p\in \mathcal{P}_k}\tilde p_u\Big(\rho_{t_k}^2\alpha_{mkjip}^{(1)}+\bar \rho_{t_k}^2\alpha_{mkjip}^{(2)}\Big)+\alpha_{mkji}^{(3)}  , &\text{if } m'=m 
			\end{cases}$
		}\\ [4pt] 
		\hline 
		{$\alpha_{mkjip}^{(1)}= \begin{cases}
				\pi_{mkjip}^{(1,1)}, &\text{if } p\neq k  \\
				\Tr\big(\overline \Rmat_{mk} \Deltamat_{mj}\overline \Rmat_{mk}\Deltamat_{mi}^H\big)  , &\text{if } p=k 
			\end{cases}$} \,\, \vline \,\,{$\alpha_{mkjip}^{(2)}=\pi_{mkjip}^{(1,1)}$}	\,	
		\vline \hspace{0pt}  {
			
			{$\alpha_{mkji}^{(3)}=\sigma_m^2\Tr\big(\Deltamat_{mj}\Deltamat_{mi}^H\overline \Rmat_{mk}\big)$} 
		} \\[4pt]
		\hline

	\end{tabular}\label{pi_terms} 
\end{table*}{}

\textbf{Calculation of $r_{lk}^{c}$, $\hat r_{lk}^{c}$ and $\tilde r_{lk}^{c}$:}
The covariances of DL common channel $r_{lk}^{c}=\text{cov}(a_{lk}^{c}[\lambda],a_{lk}^{c}[\lambda])$, the estimate $\hat r_{lk}^{c}=\text{cov}(\hat a_{lk}^{c}[\lambda],\hat a_{lk}^{c}[\lambda])$ and the estimation error $\tilde r_{lk}^{c}=\text{cov}(\tilde a_{lk}^{c}[\lambda],\tilde a_{lk}^{c}[\lambda])$  with  expressions:
{\begin{align}
		\hspace{-5pt}r_{lk}^{c}&=\sum_{m',m\in\mathcal{A}_{l}}\sum_{j,i\in\mathcal{K}_{l}}\sqrt{p_{lm'}^{c}p_{lm}^{c}}{\overline\eta_{lm'm}^{c}} (\pi_{m'mkji}-\varsigma_{m'mkji}) \nonum \\
		\hat r_{lk}^{c}&
		=(\theta_{lk}^{c})^{2}(\psi_{lk}^{c})^{-1}  \text{ and }
		\tilde r_{lk}^{c}=r_{lk}^{c}-\hat r_{lk}^{c}. \label{r_lkc}
\end{align}}     
\vspace{-15pt}                          
\subsection{DL Private Channel Estimation} \label{DL private channel estimation}                                         
The LMMSE estimate of  DL private channel $a_{lki}^{p}[\lambda]$ is~\cite{Yao2024}:
\begin{align}
	\hat a_{lki}^{p}[\lambda]&=\bar a_{lki}^{p}[\lambda]+\theta_{lki}^{p}(\psi_{lki}^{p})^{-1}\big(y_{lk}^{p}[\tilde t_k]-\bar y_{lk}^{p}[\tilde t_k]\big). \label{DL_estimate_private}
\end{align} 
The scalar  $\bar a_{lki}^{p}[\lambda]=\E\{a_{lki}^{p}[\lambda]\}=\bar a_{lki}^{p}[\lambda]=\sum_{m\in\mathcal{A}_l}\sqrt{p_{mi}^{p}\eta_{mi}^{p}}\Tr(\Lambdamat_{mki}\hat \Rmat_{mk})$ for $i\in\mathcal{P}_{k}$ and $0$ otherwise. The scalar $\bar y_{lk}^{p}[\tilde t_k]=\E\{y_{lk}^{p}[\tilde t_k]\}=\sum_{l=1}^{L}\sum_{i\in\mathcal{P}_k}a_{lki}^{p}[\tilde t_k]$ is the mean of the DL private pilot signal received at the $k$th UE, and  is given in \eqref{DL_pilot_private}. The covariance 
$\theta_{lki}^{p}=\text{cov}(y_{p,k}[\tilde t_k], a_{lki}^{p}[\lambda])$ and  $\psi_{lki}^{p}=\text{cov}(y_{p,k}[\tilde t_k],y_{p,k}[\tilde t_k])$ are  calculated on lines similar to $\theta_{lk}^{c}$ and $\psi_{lk}^{c}$. Their simplified expressions are given in Table~\ref{pi_terms}.\\
\textbf{Calculation of $r_{lkj}^{p}$, $\hat r_{lkj}^{p}$ and $\tilde r_{lk}^{p}$}:
The covariance of DL private channel $r_{lkj}^{p}=\text{cov}(a_{lkj}^{p}[\lambda],a_{lkj}^{p}[\lambda])$, its estimate $\hat r_{lkj}^{p}=\text{cov}(\hat a_{lkj}^{p}[\lambda],\hat a_{lkj}^{p}[\lambda])$, and estimation error $\tilde r_{lkj}^{p}=\text{cov}(\tilde a_{lkj}^{p}[\lambda],\tilde a_{lkj}^{p}[\lambda])$ are calculated in \cite{RSMA_Tech_report}. The covariance, similar to  \eqref{r_lkc}, are given as 
\begin{align}
	\hspace{-6pt}r_{lkj}^{p}
	&=\sum_{m',m\in\mathcal{A}_{l}}\sqrt{p_{m'j}^{p}p_{mi}^{p}}{\overline \eta_{m'mji}^{p}}(\pi_{m'mkji}-\varsigma_{m'mkji}) \nonum \\
	\hspace{-6pt}\hat r_{lkj}^{p}&=\theta_{lki}^{p}(\psi_{lki}^{p})^{-1}\theta_{lki}^{p} \text{  and } \hspace{2pt} \tilde r_{lkj}^{p}=r_{lkj}^{p}-\hat r_{lkj}^{p}. \label{r_hat_kj}
\end{align}

\bibliographystyle{IEEEtran}
\bibliography{IEEEabrv,ChannelAging_References}
\end{document}